\documentclass[pra,twocolumn,superscriptaddress,showpacs]{revtex4}
\usepackage{bbm}
\usepackage{amsmath, amssymb}
\usepackage{color}
\usepackage{graphicx,epsfig}
\usepackage{pifont}
\usepackage{subfigure}
\usepackage{amsmath, amssymb, graphics}

\begin{document}

\title{Dissipation assisted quantum correlations in coupled qubits}
\author{Chaitanya Joshi}
\email{chaitanya.joshi@york.ac.uk}
\affiliation{Department of Physics and York Centre for Quantum Technologies, University of York, Heslington, York, YO10 5DD, UK }
\date{\today}

\begin{abstract}
We theoretically investigate a possibility to establish  multi-qubit quantum correlations in one-dimensional chains of qubits. We combine a reservoir engineering strategy  with coherent dynamics to generate multi-qubit entangled states. We find that an interplay between the coherent and incoherent dynamics result in the generation of stable (time-independent) many-body entangled steady states. Our results will be relevant in the context of the dissipative generation of quantum states, with applications in short-distance quantum computation and for exploring the emergence of collective phenomena in many-body open quantum systems.
\end{abstract}

\pacs{42.50.-p,03.67.-a,75.10.Pq}
\maketitle
\section{Introduction}
Quantum technology promises novel techniques to process information at a level and scale which is inconceivable in the classical domain \cite{aste98,chbe00}. Technological  advancements in the past  two decades  have fueled the transformation of early theoretical ideas and protocols highlighting quantum weirdness into experimental reality \cite{shar13,djwi13}. Over the years, different quantum systems have received  growing attention for exploring them in various roles, including quantum simulators \cite{rfey82,dporr04,kkim11,rbla12,imge14} and  processors for various tasks in quantum computation and information \cite{tdla10}. Unfortunately, almost every quantum system, that is of potential interest to us, also irreversibly  couples to its external environment. This normally leads to decoherence and dissipation in quantum systems \cite{whzu03}. It is, therefore, advisable to minimize the influence of reservoir induced decoherence on a quantum system. In this direction, quantum control strategies have taken a center stage for quantum state protection in noisy quantum systems 
(see \cite{arrrc01,bmte15,csay11,kfui14,jkel15,sjgl15,yliu16,cjosh16}, and references therein). 

Somewhat counterintuitively, it is also possible to engineer the irreversible system-reservoir coupling in order to prepare desired many-body quantum states \cite{sdie08,fver09,hkra11,jtba11,ylin13,tram14}.
These schemes are based on the method of {\it reservoir engineering} \cite{jfpo96} to tailor the system-reservoir coupling and drive the quantum system to a desired quantum state. In this work, we follow one such reservoir engineering strategy considered  in \cite{sdie08,jtba11} and combine it with coherent interactions  to create stable many-body quantum states. Prime motivation behind our work is our shared belief that the influence of environment induced dissipation can also be particularly  intriguing when the quantum system of interest itself is already an interacting many-body system. In such a scenario, an interesting interplay between the coherent and incoherent interactions  can result in the generation of non-trivial many-body states with applications in quantum technologies \cite{sdie08,fver09,hkra11,jtba11,ylin13,tram14,cchen16}.

In this work, we  will specifically focus on one-dimensional chains of two-level systems (or, ``qubits'') to engineer stable quantum many-body states. The prospects of using qubit chains as ``non-photonic'' alternates   for short distance quantum communication  has already received considerable attention in the past \cite{sbos03,sbos07,akay10}.  These strategies exploit the inter-qubit coherent interactions to accomplish some of the necessary tasks in quantum computation and information processing. Specific applications include ``quantum interconnects'' to join two or more quantum processors, ``quantum channels'' for quantum state transfer and entanglement generation between distant qubits \cite{sbos03,sbos07,akay10,idam07,mchr04,lami08,tscu08,cdfr08,nyya11,lban11,fcar14,ssah15,rjch16}. In contrast to previous proposals, our current agenda will be to combine the ideas from reservoir engineering with coherent dynamics between the qubits to create  time-independent many-body quantum states. Specifically, we will use the method of reservoir engineering to create a two-qubit entangled state. This dissipatively generated entangled state, evolving under competing coherent interactions, transforms  to yield a multi-qubit entangled state.

The open dynamics of an interacting many-body open quantum system  can often be well described by a Lindblad type master equation under the Born-Markov and secular approximations \cite{hpbr02},
\begin{equation}\label{open}
\dot{\hat{\rho}}= -i[\hat H_{\rm S},\hat{\rho}]+\Gamma\hat{\mathcal L}_{\hat A}\hat{\rho},
 \end{equation}
where $\hat\rho$ is the system's state, $\hat H_{\rm S}$ is the system Hamiltonian, and $\hat{\mathcal L}_{\hat A}\hat{\rho}$ is the Lindblad super operator rendering the effects of the reservoir on the system ($\hat A$ is the so called quantum jump operator).  The steady state is solved by letting $i[\hat H_{\rm S},\hat{\rho}_\mathrm{ss}]-\Gamma\hat{\mathcal L}_{\hat A}\hat{\rho}_\mathrm{ss}=0$. In the sections to follow next, 
 we will show that a carefully designed reservoir coupling when combined with coherent evolution  can indeed result in such stable (time-independent) many-body quantum states.

 \section{Model: two qubits} We begin with considering  a physical system composed of two qubits with their closed dynamics governed by a Hamiltonian of the form  ($\hbar=1$), 
\begin{equation}\label{systwoqubit}
\hat H_{\rm P}=\Delta(\hat\sigma_1^{z}+\hat\sigma_2^{z}),
\end{equation}
where $\Delta$ is the (identical) energy level splitting between the two energy levels of each qubit.  Throughout in this paper we will use dimensionless parameters such that we will scale all frequencies by $\Delta$ and time by $\Delta^{-1}$. Nevertheless, we will keep $\Delta$ in all expressions and will always use $\Delta=1$ in any calculations. For reasons that will soon become clear, the two qubits evolving under the Hamiltonian \eqref{systwoqubit} form what we call the {\it primary} chain, and this is reflected in our choice of notation $\hat H_{\rm P}$ in equation \eqref{systwoqubit}. We start with revisiting a reservoir engineering strategy originally proposed in \cite{sdie08} and experimentally realized in a linear ion-trap quantum computer architecture in \cite{jtba11}.  We, thus, envision an open version scenario where  the two qubits in the primary chain are also coupled to an engineered reservoir, 
\begin{equation}\label{open2qubitorig}
\dot{\hat{\rho}}= -i[\hat H_{\rm P},\hat{\rho}]+\Gamma\hat{\mathcal L}_{{\hat b}_{1,2}}\hat{\rho},
 \end{equation}
 where  ${\hat b}_{1,2}=(\hat\sigma_1^{+}+\hat\sigma_2^{+})(\hat\sigma_1^{-}-\hat\sigma_2^{-})$ is the bi-local quantum jump operator \cite{sdie08,jtba11}. As discussed in detail in Refs.~\cite{sdie08,jtba11}, the above quantum jump operator map any antisymmetric component in the wavefunction on a pair of qubits into the symmetric one. We refer the reader to \cite{jtba11} for a physical realization of the above quantum jump operator in trapped atomic ions. In order to understand the role of the bi-local quantum jump operator in the above master equation \eqref{open2qubitorig}, we assume that the two qubits are initialized in a pure state $|\psi(0)\rangle=|\uparrow\rangle_{1}|\downarrow\rangle_{2}$ with zero angular momentum projection along the $z$ direction. The bi-local operator $ {\hat b}_{1,2}$ acting on this initial gives  ${\hat b}_{1,2}|\uparrow\rangle_{1}|\downarrow\rangle_{2}=|\uparrow\rangle_{1}|\downarrow\rangle_{2}+|\downarrow\rangle_{1}|\uparrow\rangle_{2}=|\Psi^{+}\rangle$. The projector $|\Psi^{+}\rangle \langle \Psi^{+}|$ also commutes with the two-qubit Hamiltonian \eqref{systwoqubit} and, therefore, the two qubits evolving under the master equation \eqref{open2qubitorig} reach a pure steady state $\rho_{\rm} \sim |\Psi^{+}\rangle \langle \Psi^{+}|$. It is worth pointing out that in the absence of non-zero direct coupling between the qubits in the primary chain, the creation of a maximally entangled pure state $|\Psi^{+}\rangle$ crucially depends on the initial state $|\psi(0)\rangle$. Symmetric initial states $|\downarrow\rangle_{1}|\downarrow\rangle_{2},|\uparrow\rangle_{1}|\uparrow\rangle_{2}$ are the stationary states of the master equation \eqref{open2qubitorig}.
%
%
%
%
%
 \begin{figure}[t]
  \centering
       \includegraphics[width=0.49\textwidth]{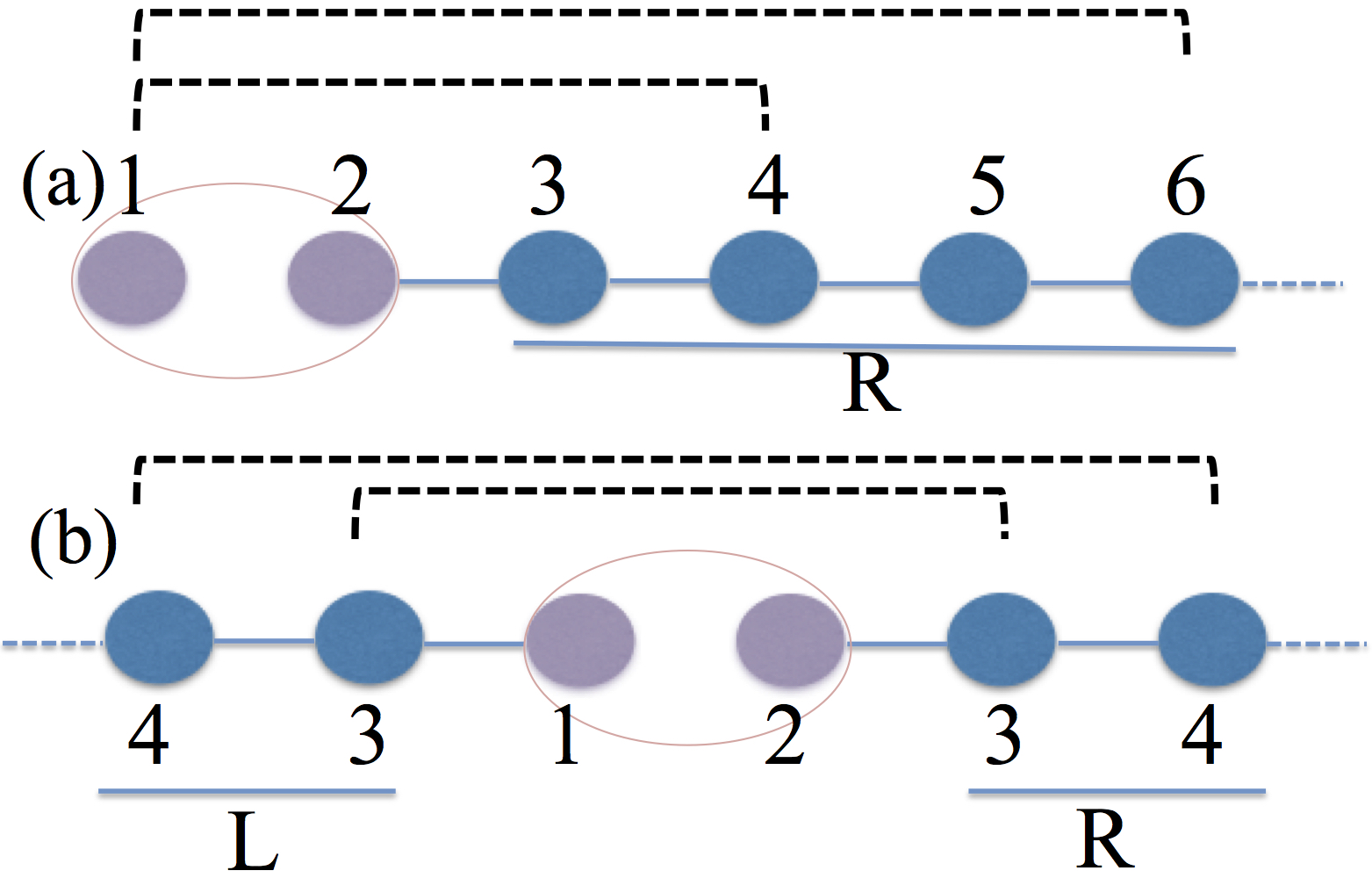}
        \caption{(Color online) ({\rm a}) The primary two-qubit chain  is coupled  to the  secondary chain of qubits on its ``right''. The goal is to prepare a many-body quantum state exhibiting steady state correlations  between the 1$^{\rm st}$  and $n^{\rm th}$ qubits (joined in dashed). ({\rm b})  The primary two-qubit chain is situated in the ``center'' and is  connected to the two secondary chains on its ``left'' and ``right'' respectively. The goal is to establish bi-partite correlations between the qubits placed at the remote (open) ends of the two secondary chains (joined in dashed). In both these configurations the primary qubits are assumed to couple to a common engineered reservoir, with no direct coupling between them.}\label{spinchainstopology}
\end{figure}

\section{Model: many qubits}
We now consider physical  scenarios where additional ({\it  secondary}) chains of coupled qubits  are joined with the two-qubit  primary chain in two different spatial configurations, as shown in Fig.\ref{spinchainstopology}.  In what follows next, we will explore the open dynamics of the coupled qubits of the primary and secondary chains arranged in these two different geometries. 
 \subsection{Qubit chain: geometry {\rm(a)}}
 We start with describing the evolution of the qubits in the primary and secondary chains, arranged as shown in Fig.\ref{spinchainstopology}(a). We model the closed dynamics of these coupled qubits 
 under the following Hamiltonian, 
\begin{equation}\label{manyqubits} 
\hat H_{\rm P-S}^{\rm a}=\hat H_{\rm P}+\hat H_{\rm S}^{\rm{R}} +\hat H_{\rm cplng}^{\rm P-R}, 
\end{equation}
where $\hat H_{\rm P}$, given by  equation \eqref{systwoqubit}, describes the evolution of the two qubits in the primary chain. The secondary chain which is coupled to the primary chain on its ``right''  is modeled  as a collection of coupled qubits under a Hamiltonian, 
\begin{equation}
\hat H_{\rm S}^{\rm R}=\sum_{j=3,4...}\Delta\hat\sigma_{j,{\rm R}}^{z}+\kappa(\hat\sigma_{j,\rm{R}}^{+}\hat\sigma_{j+1,\rm{R}}^{-}+h.c.),
 \end{equation}
 where $\hat\sigma_{j,{\rm R}}^{z}$ represents the Pauli-$z$ operator of the $j^{\rm th}$ qubit in the secondary chain and the inter-qubit coupling strength is denoted by $\kappa$. The coherent coupling between the primary and secondary chains is assumed to take a form,
 \begin{equation}
\hat H_{\rm cplng}^{\rm P-R}= \theta(\hat\sigma_2^{+}\hat\sigma_{3,\rm{R}}^{-}+h.c),
 \end{equation}
 where  $\theta$ is the inter-chain coupling strength.  In the discussion to follow next, we will fix $\theta=\kappa$.
  
 We now use a following strategy to achieve  quantum many-body states of qubits in the primary and secondary chains. We assume that the qubits in the secondary chain remain decoupled from the external surroundings and the qubits in the primary chain couple to an engineered  reservoir through the action of the bi-local quantum jump operator ${\hat b}_{1,2}$. The qubits in the secondary chain can be protected from external noisy environment through a quantum control strategy such as dynamical decoupling. The idea behind dynamical decoupling is to rapidly rotate the quantum system by means of classical fields in order to average the system-environment coupling to zero \cite{gdel10,aber12,nbar13}. The collective open dynamics of the qubits in the primary and secondary chains can then be modeled   by a master equation of the form, 
  \begin{equation}\label{openmeqn}
\dot{\hat{\rho}}= -i[\hat H_{\rm P-S}^{\rm a},\hat{\rho}]+\Gamma\hat{\mathcal L}_{{\hat b}_{1,2}}\hat{\rho}.
 \end{equation}
 
 It is easy to verify that the evolution under the master equation \eqref{openmeqn} conserves the initial number of excitations present in the  primary and secondary chains. We, therefore, initialize the qubits in a non-vacuum state. Specifically, we  assume that  the qubits in the primary chain are initially in a state $|\psi(0)\rangle=|\uparrow\rangle_{1}|\downarrow\rangle_{2}$ and all qubits in the secondary chain are initialized in their ground states  $|\psi_{j,R}(0)\rangle=|\downarrow\rangle_{j}~(j=3,4...)$. Our claim is, this initial state, evolving under the  master equation \eqref{openmeqn},  will evolve to a pure many-body  entangled state of the primary and secondary qubits. In order to show this, we first numerically simulate  the master equation \eqref{openmeqn} and then provide an analytical explanation for the results. We simulate the master equation \eqref{openmeqn} for even number of qubits ($2N$) in the secondary chain. Therefore, the  total number of qubits in  the primary and secondary chains is $n=2+2N~(N=1,2,3...)$. We use the method of  time evolving block decimation (TEBD) extended to open systems (mixed states) \cite{mzwo04} to numerically simulate our many-body master equation \eqref{openmeqn}. This numerical method has been previously applied to explore non-equilibrium features in driven-dissipative many-body quantum systems \cite{cjosh13}. Our numerical approach allows us to compute single- and double- site correlators, which is suffice to reconstruct the reduced density matrix of  
 any two qubits in the primary and secondary chains. 
 
\begin{figure}[h!]
  \centering
    \includegraphics[width=0.49\textwidth]{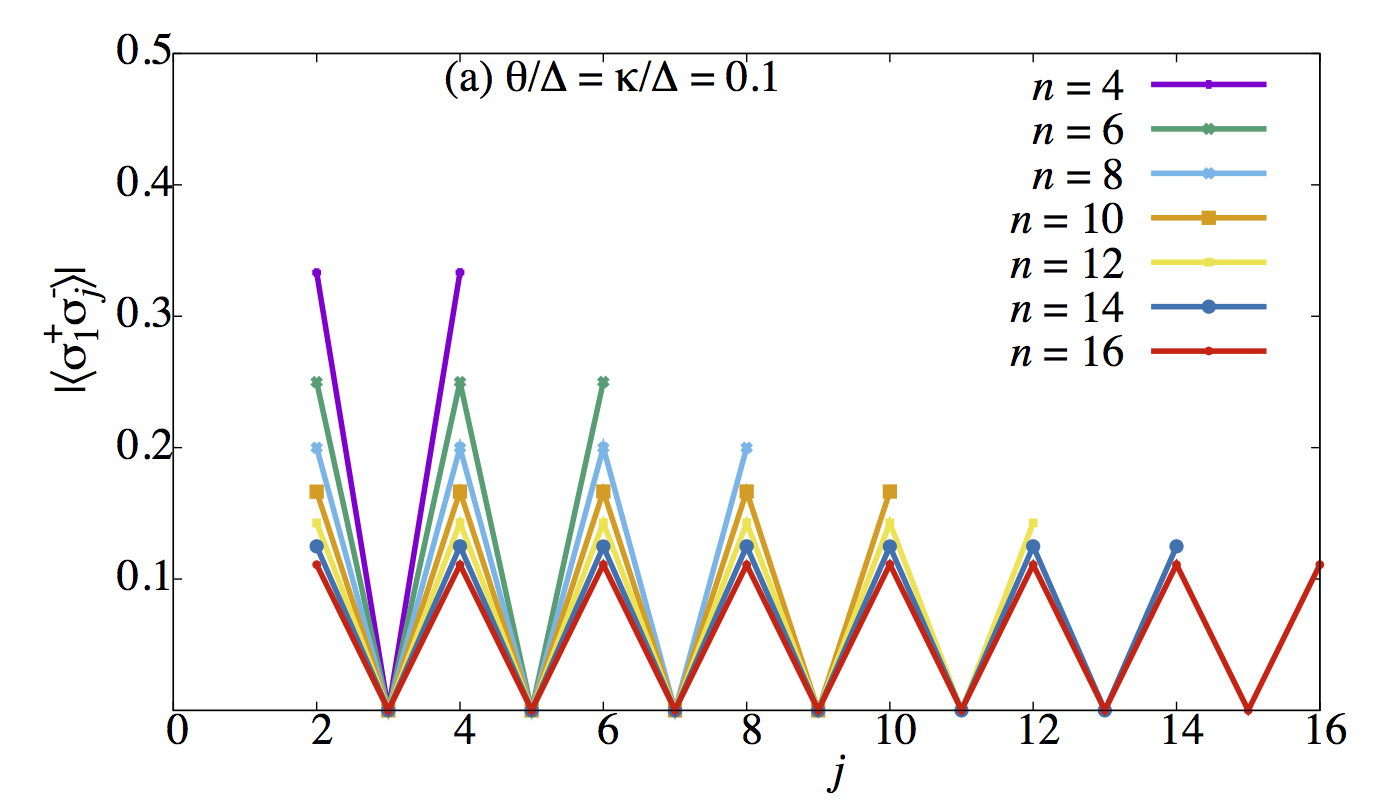}\\
    \includegraphics[width=0.49\textwidth]{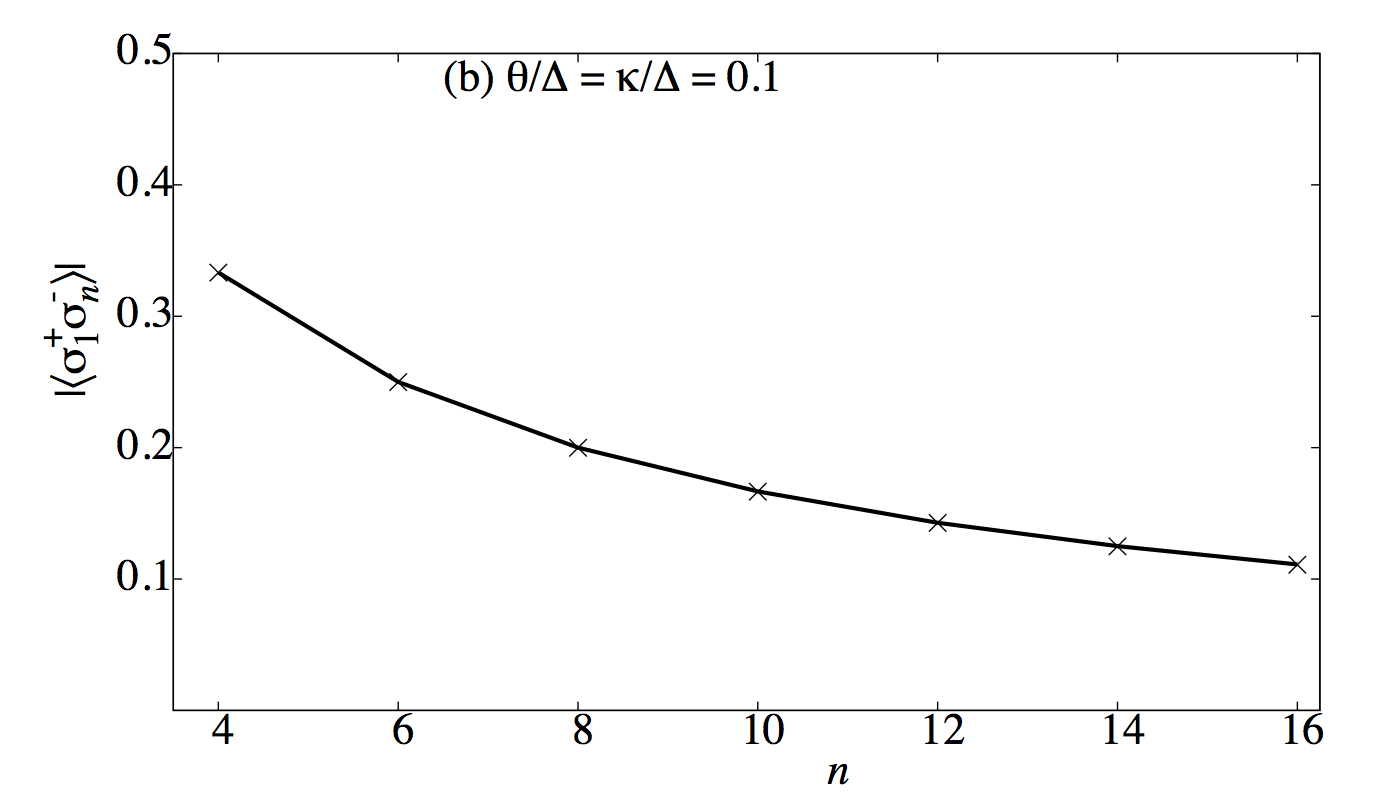}\\
     \includegraphics[width=0.48\textwidth]{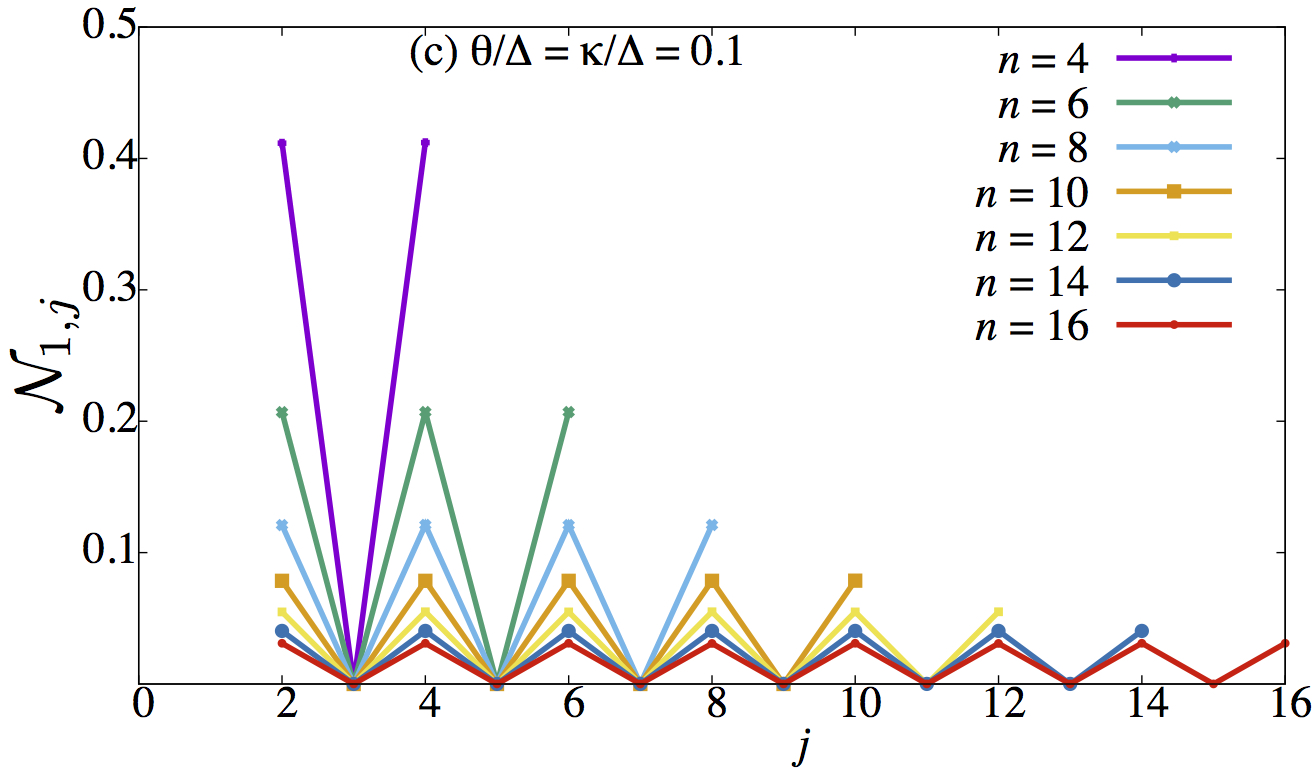}
        \caption{(Color online){\rm(a)} Steady state value of the correlator $|\langle \hat{\sigma}_{1}^{+}\hat{\sigma}_{j}^{-}\rangle|$ capturing the pairwise correlations between the first qubit in the primary chain and all other qubits arranged as shown in Fig.\ref{spinchainstopology}(a). {\rm(b)} Steady state value of the correlator $|\langle \hat{\sigma}_{1}^{+}\hat{\sigma}_{n}^{-}\rangle|$ capturing the correlations between the 1$^{\rm st}$ and $n^{\rm th}$ qubits. The numerical results (dot) match perfectly with the analytical result (solid line), confirming  linear decay $1/(n/2+1)$ of  correlations. {\rm(c)} Steady state value of the negativity $\mathcal N_{1,j}$ capturing the degree of pairwise entanglement  between the first qubit in the primary chain and all other qubits arranged in a configuration shown in Fig.\ref{spinchainstopology}(a).The other dimensionless parameters are $\Delta=1$ and $\Gamma=0.1$.}\label{diffqubits}
\end{figure}

 In Fig.\ref{diffqubits}(a), we plot the steady state correlator $|\langle \hat{\sigma}_{1}^{+}\hat{\sigma}_{j}^{-} \rangle|$ as a function of index $j$. This captures the degree of pairwise correlations present between the first qubit in the primary chain and all other qubits, arranged as shown in Fig.\ref{spinchainstopology}(a). We have now dropped the subscript ``R'' in the Pauli raising and lowering operators for the qubits in the secondary chain. One interesting observation from Fig.\ref{diffqubits}(a) is that the steady state correlator  $|\langle \hat{\sigma}_{1}^{+}\hat{\sigma}_{j}^{-} \rangle|$ for coupled qubits' arrangement of Fig.\ref{spinchainstopology}(a)  follow a pattern, 
 \begin{eqnarray*}
 |\langle \hat{\sigma}_{1}^{+}\hat{\sigma}_{j}^{-} \rangle|&=& 0, {\rm when}~j \in {\rm odd},\\
  |\langle \hat{\sigma}_{1}^{+}\hat{\sigma}_{j}^{-} \rangle|&=& {\rm constant}, {\rm when}~j \in {\rm even},
 \end{eqnarray*}
 where the value of the constant depends only on the total number of qubits $n$ in the chain and is completely independent of $j$. Furthermore, from the numerical solution of the  master equation \eqref{openmeqn} we find that $\langle \hat{\sigma}_{j}^{z}\rangle = -1~\forall~j \in {\rm odd}~(j=3,5,7...)$, while $\langle \hat{\sigma}_{j}^{z}\rangle~(j=1,2,4...)$ depends on the total number of qubits $n$ in the chain (see below). The TEBD method provides an efficient way to simulate our many-body master equation \eqref{openmeqn}, but we do not have access to the full steady state density matrix itself. Therefore, we  numerically diagonalize  the master equation \eqref{openmeqn} for small values of $n$ to gain more insight into the steady state of the master equation \eqref{openmeqn} and the correlation pattern of Fig.\ref{diffqubits}(a). For example, an exact numerical diagonalization of the master equation \eqref{openmeqn} for $n=4$ result in a following pure steady state,
 \begin{eqnarray}\label{stdyfrn4}
 |\Psi_{\rm n=4}\rangle&=&\frac{1}{\sqrt{3}}( |\uparrow \rangle_{1}|\downarrow \rangle_{2}|\downarrow \rangle_{4}+|\downarrow \rangle_{1}|\uparrow \rangle_{2}|\downarrow \rangle_{4}\nonumber \\
 && -|\downarrow \rangle_{1}|\downarrow \rangle_{2}|\uparrow \rangle_{4})\otimes |\downarrow \rangle_{3}.
 \end{eqnarray}
 
It is easy to verify that $ |\Psi_{\rm n=4}\rangle$ is an eigenstate of the Hamiltonian $\hat H_{\rm P-S}^{\rm a}$ \eqref{manyqubits} and, thus, the corresponding  steady state $\hat{\rho}_{\rm n=4}~(|\Psi_{\rm n=4}\rangle \langle \Psi_{\rm n=4}|)$ commutes with the Hamiltonian $\hat H_{\rm P-S}^{\rm a}$. Moreover, the state $\hat{\rho}_{\rm n=4}$ is also a symmetric state of the qubits in the primary chain (qubits 1 and 2) and, therefore, also a {\it dark state} of the bi-local Lindblad operators $\hat{\mathcal L}_{{\hat b}_{1,2}}\hat{\rho}_{\rm n=4}$. It is quite remarkable to observe that the steady state solutions $\hat{\rho}_{\rm n=4}$ is independent of the inter-qubit and inter-chain coupling strengths. Of course, these coupling strength determine the time scale for the production of the steady state $\hat{\rho}_{\rm n=4}$. It is immediately clear that the state \eqref{stdyfrn4} is similar to a three-qubit $W$ state, except for the weighting factors. Such a $W$ state represent one of the two entangled  classes of three-qubit states with several applications in quantum information theory, the other being the $GHZ$ states \cite{wdur00,vngo06,bookchuang}. $W$ states have interesting properties, including a non-zero bi-partite entanglement between any pair of qubits and robustness of quantum correlations  against loss of one qubit. Likewise, through a direct numerical diagonalization of the master equation \eqref{openmeqn} we obtain following pure steady states, 
 \begin{eqnarray}\label{stdyfrn6and8}
 |\Psi_{\rm n=6}\rangle&=&\frac{1}{\sqrt{4}}( |\uparrow \rangle_{1}|\downarrow \rangle_{2}|\downarrow \rangle_{4}|\downarrow \rangle_{6}+|\downarrow \rangle_{1}|\uparrow \rangle_{2}|\downarrow \rangle_{4}|\downarrow \rangle_{6} \nonumber \\
 &&-|\downarrow \rangle_{1}|\downarrow \rangle_{2}|\uparrow \rangle_{4}|\downarrow \rangle_{6}+|\downarrow \rangle_{1}|\downarrow \rangle_{2}|\downarrow \rangle_{4}|\uparrow \rangle_{6}) \nonumber \\
 &&\otimes 
  |\downarrow \rangle_{3} |\downarrow \rangle_{5},\\  \nonumber
 |\Psi_{\rm n=8}\rangle&=&\frac{1}{\sqrt{5}}( |\uparrow \rangle_{1}|\downarrow \rangle_{2}|\downarrow \rangle_{4}|\downarrow \rangle_{6}|\downarrow \rangle_{8}+|\downarrow \rangle_{1}|\uparrow \rangle_{2}|\downarrow \rangle_{4}\nonumber \\
 &&|\downarrow \rangle_{6}|\downarrow \rangle_{8}
 -|\downarrow \rangle_{1}|\downarrow \rangle_{2}|\uparrow \rangle_{4}|\downarrow \rangle_{6}|\downarrow \rangle_{8}+|\downarrow \rangle_{1}|\downarrow \rangle_{2}\nonumber \\
 &&|\downarrow \rangle_{4}|\uparrow \rangle_{6}|\downarrow \rangle_{8}
 -|\downarrow \rangle_{1}|\downarrow \rangle_{2}|\downarrow \rangle_{4}|\downarrow \rangle_{6}|\uparrow \rangle_{8})\nonumber \\
 &&\otimes 
  |\downarrow \rangle_{3} |\downarrow \rangle_{5} |\downarrow \rangle_{7}.\\  \nonumber
 \end{eqnarray}
As expected, the above are eigenstates of the Hamiltonian $\hat H_{\rm{P-S}}^{\rm a}$ \eqref{manyqubits} and also  dark states of the bi-local Lindblad operators.  The structure of the steady states $|\Psi_{\rm n=4,6,8}\rangle$ clearly explains the numerical features observed in Fig.\ref{diffqubits}(a).

It is easy to generalize the above  to obtain an exact expression for the steady state of the master equation \eqref{openmeqn} for arbitrary large (even) number of qubits in the primary and secondary chains. 
In general, the steady state of the master equation \eqref{openmeqn} will be an entangled  $W$ state of $n/2+1$ qubits, with rest of the $n/2-1$ ``odd-numbered'' secondary qubits 
in their respective ground states. The value of the correlator $|\langle \hat{\sigma}_{1}^{+}\hat{\sigma}_{j}^{-} \rangle|$
 is independent of the index $j$ and is solely dependent on the total number of qubits in the $W$ state and, thus, scale as $\sim 1/(n/2+1)$. In Fig.\ref{diffqubits}(b) we have shown the steady state correlator $|\langle \hat{\sigma}_{1}^{+}\hat{\sigma}_{n}^{-} \rangle|$ capturing the correlations between  the $1^{\rm st}$ qubit in the primary chain and the remote $n^{\rm th}$ qubit of the secondary chain. 
 The numerical results (dot) match perfectly with the analytical result (solid line) confirming linear decay of  correlations $\sim 1/(n/2+1)$. In order to quantify the degree of true quantum correlations present between  the first qubit in the primary chain and all other qubits arranged in a configuration shown in Fig.\ref{spinchainstopology}(a), we will use negativity $\mathcal N_{1,j}$ defined as
 \begin{equation}
 \mathcal N_{1,j}= max (0,\sum_{k=1}^{4}|\lambda_{k}|-1),
 \end{equation}
where $\lambda_{k}$ are the eigenvalues of the partially transposed two-qubit density matrix $\rho_{1,j}^{T_{j}}$ \cite{bookchuang}. As expected, the resulting pairwise entanglement shown in Fig.\ref{diffqubits}(c) also follow a pattern similar to Fig.\ref{diffqubits}(a).To conclude, in this section we have provided a scheme to {\it transform} a locally prepared two-qubit maximally entangled state to a multi-qubit $W$ state with multi-partite entanglement. We have combined a reservoir engineering with coherent interactions to accomplish  this task. In the past, multi-party $W$ states have been experimentally generated using carefully controlled pulse sequences in different physical systems, including trapped ions \cite{mhaf05} and superconducting qubits \cite{nmat10}. We also refer the reader to \cite{frei15,caro16}, for alternate proposals on the dissipative engineering of multi-party $W$ states. Dissipation-induced correlations has also been investigated in one-dimensional  systems of interacting bosons \cite{mkif11}. 

 \subsection{Qubit chain: geometry \rm{(b)}}
We now consider another  physical scenario  of practical interest, where the primary chain  is coupled to the two identical  secondary chains on its ``left'' and ``right'', as shown in  Fig.\ref{spinchainstopology}(b). The goal is to establish bi-partite correlations between the qubits placed at the remote (open) ends of the two secondary chains. States generated through links joining the qubits located symmetrically with respect to the center (as shown in Fig.\ref{spinchainstopology}(b)) are also known as rainbow states and have been previously  explored in detail in \cite{gvit10, gram14}. As before, we model the closed dynamics of the qubits in the primary and secondary chains  under the following Hamiltonian, 
\begin{equation}\label{manyqubitssetup2ham} 
\hat H_{\rm P-S}^{\rm b}=\hat H_{\rm S}^{\rm{L}}+\hat H_{\rm P}+\hat H_{\rm S}^{R} +\hat H_{\rm cplng}^{\rm P-L}+\hat H_{\rm cplng}^{\rm P-R}, 
\end{equation}
where $\hat H_{\rm P}$ describes the free evolution of the qubits in the primary chain given by equation \eqref{systwoqubit} and,
\begin{eqnarray*}
\hat H_{\rm S}^{\rm L (R)}&=&\sum_{j=3,4...} \Delta \hat\sigma_{j,\rm{L (R)}}^{z}+\kappa(\hat\sigma_{j,\rm{L (R)}}^{+}\hat\sigma_{j+1,\rm{L (R)}}^{-}+h.c.),~~~~\\
\hat H_{\rm cplng}^{\rm P-L (R)}&=& \theta(\hat\sigma_{3,\rm{L (R)}}^{+}\hat\sigma_{1 (2)}^{-}+h.c.).
 \end{eqnarray*}
As done in the previous section, we model the open dynamics of the qubits' arrangement of  Fig.\ref{spinchainstopology}(b) under a master equation of the form,
 \begin{equation}\label{open2qubit}
\dot{\hat{\rho}}= -i[\hat H_{\rm P-S}^{\rm b},\hat{\rho}]+\Gamma\hat{\mathcal L}_{{\hat b}_{1,2}}\hat{\rho},
 \end{equation}
 where we have again assumed that the qubits in the primary chain are coupled to an engineered reservoir through the action of the bi-local quantum jump operator ${\hat b}_{1,2}$ and the qubits in the secondary chains are assumed to remain noise-free. We assume that the qubits in the primary chain are initially in a state $|\psi(0)\rangle=|\uparrow\rangle_{1}|\downarrow\rangle_{2}$ and the qubits in the left and  right secondary chains are initialized in their ground states  $|\psi_{j,\rm{L(R)}}(0)\rangle=|\downarrow\rangle_{j,\rm{L(R)}}~(j=3,4...)$. We again use the TEBD numerical method for mixed states to time evolve the master equation \eqref{open2qubit}. We find that the time-evolved density matrix $\hat{\rho}(t)$ is a many-body entangled state of the qubits, which exhibit bi-partite quantum correlations across all partitions of the qubits in the  primary and secondary chains. However, in contrast to the previous section, numerically obtained $\hat{\rho}(t)$ is a time-dependent mixed state. Furthermore, numerical simulation of the  master equation \eqref{open2qubit} suggests that the time-evolved state $\hat{\rho}(t)$ becomes more mixed as the inter-qubit coupling strengths $\kappa,\theta$ increases.
 
In order to understand  the structure of the time-evolved density matrix $\hat{\rho}(t)$, it is suffice to restrict ourselves to a case when $n$ = 4. Since the master equation \eqref{open2qubit} conserves the initial number of excitations, the states $| \uparrow\rangle_{3,\rm{L}}| \downarrow\rangle_{1}|\downarrow\rangle_{2}|\downarrow\rangle_{3,\rm{R}}; | \downarrow\rangle_{3,\rm{L}}| \uparrow\rangle_{1}|\downarrow\rangle_{2}|\downarrow\rangle_{3,\rm{R}}; | \downarrow\rangle_{3,\rm{L}}| \downarrow\rangle_{1}|\uparrow\rangle_{2}|\downarrow\rangle_{3,\rm{R}}; | \downarrow\rangle_{3,\rm{L}}| \downarrow\rangle_{1}|\downarrow\rangle_{2}|\uparrow\rangle_{3,\rm{R}}$ form a complete basis set. In this subspace, the Hamiltonian \eqref{manyqubitssetup2ham} has two eigenstates which are symmetric states of the 
two primary qubits  ($\theta=\kappa$),
\begin{eqnarray}
|\Phi^{\pm}\rangle&=&\frac{1}{2}( \mp(|\uparrow \rangle_{3,\rm{L}}|\downarrow \rangle_{3,\rm{R}}+|\downarrow \rangle_{3,\rm{L}}|\uparrow \rangle_{3,\rm{R}})|\downarrow \rangle_{1}|\downarrow \rangle_{2} \nonumber \\ 
&&+(|\uparrow \rangle_{1}|\downarrow \rangle_{2}+|\downarrow \rangle_{1}|\uparrow \rangle_{2})|\downarrow \rangle_{3,\rm{L}}|\downarrow \rangle_{3,\rm{R}} \nonumber),
\end{eqnarray}
with eigenenergies $-(2\Delta \pm \kappa)$. Both the states $|\Phi^{\pm}\rangle$ are dark states of the bi-local Lindblad operator and, therefore, $\hat{\rho}(t)$ is a time-dependent mixture of the states $|\Phi^{\pm}\rangle$. This argument  can be easily extended to larger values of $n$ to understand the structure of the time-evolved state $\hat{\rho}(t)$.

\begin{figure}[h!]
  \centering
    \includegraphics[width=0.49\textwidth]{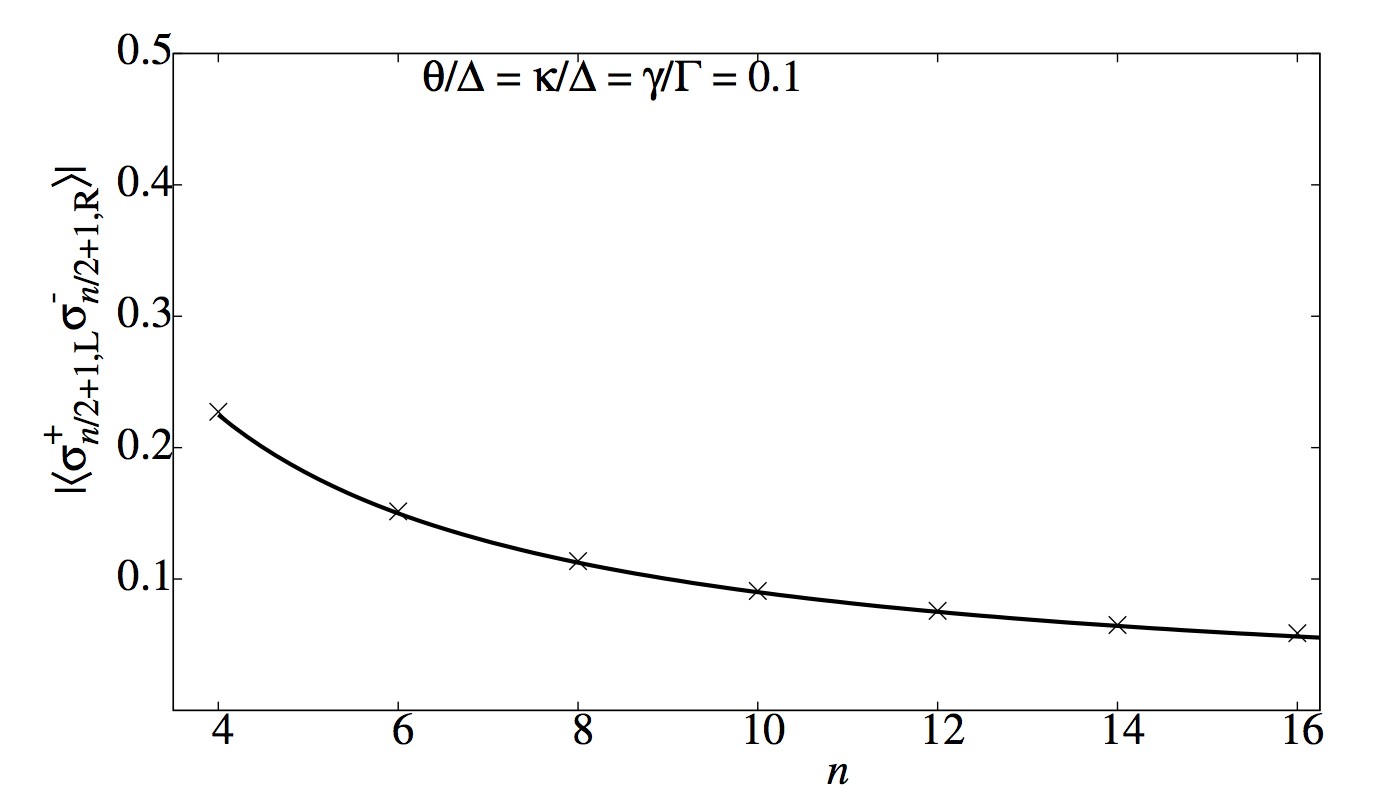}\\
     \includegraphics[width=0.46\textwidth]{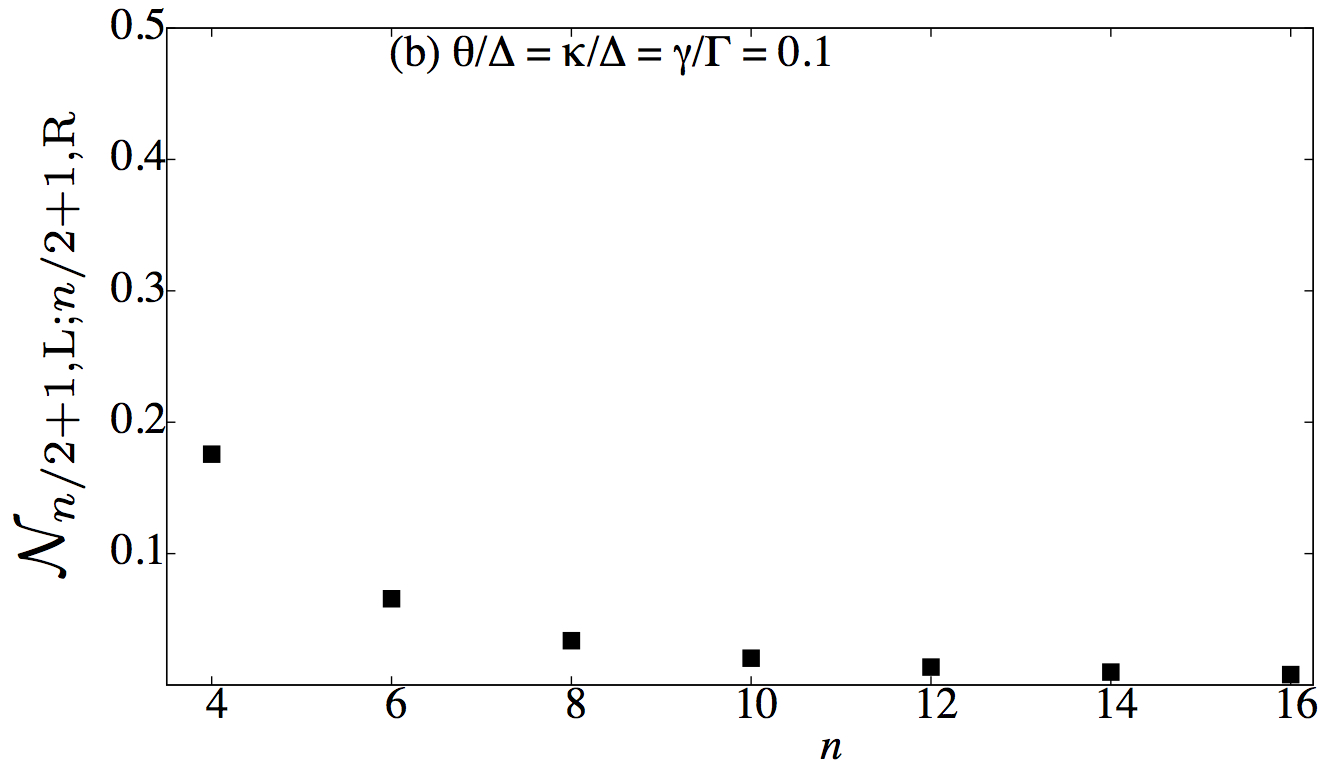}
        \caption{(Color online) {\rm(a)} The steady state correlator $\langle \hat{\sigma}_{n/2+1,\rm{L}}^{+}\hat{\sigma}_{n/2+1,\rm{R}}^{-}\rangle|$, and {\rm(b)} the pairwise quantum entanglement  $\mathcal N_{n/2+1,{\rm L};n/2+1,{\rm R}}$ shown as a function of $n$, capturing the correlations between the  qubits placed at the remote (open) ends of the two secondary chains 
coupled to the two-qubit primary chain, as illustrated in Fig.\ref{spinchainstopology}(b). The numerical results (dot) confirm a linear drop of  correlations $\sim1/n$ (solid line) in Fig.\ref{spinchainstopology}(a). The other dimensionless parameters are $\Delta=1$ and $\Gamma=0.1$ }\label{diffqubitssetup2}
\end{figure}

In order to stabilize the time-dependent fluctuations in the dynamics of the qubits evolving under the master equation \eqref{open2qubit}, we  propose to couple the qubits in the primary chain  to two independent dephasing baths. As we will show below, such a counterintuitive arrangement is capable of achieving a time-independent (mixed) entangled steady state  of the qubits in the primary and secondary chains. In the presence of additional dephasing baths,  the joint dynamics of the qubits in the primary and secondary chains can be modeled as, 
\begin{equation}\label{open2qubitdep}
\dot{\hat{\rho}}= -i[\hat H_{\rm P-S}^{\rm b},\hat{\rho}]+\Gamma\hat{\mathcal L}_{{\hat b}_{1,2}}\hat{\rho}+\gamma\hat{\mathcal L}_{{\hat \sigma}_{1}^{z}}\hat{\rho}+\gamma\hat{\mathcal L}_{{\hat \sigma}_{2}^{z}}\hat{\rho}.
 \end{equation}
We find that  a finite but non-zero dephasing rate $\gamma$ can result in a time-independent steady state which is a mixture of bi-partite entangled states.  For instance, an exact numerical diagonalization of the master equation \eqref{open2qubitdep} for $n=4$ suggests that the steady state is a mixture of entangled states $\hat{\rho}_{1;2}$ and  $\hat{\rho}_{3,\rm{L};3,\rm{R}}$. We next use the TEBD method to solve for the steady state of the master equation \eqref{open2qubitdep} for different values of $n$. Our numerical simulation of the master equation \eqref{open2qubitdep} corroborates  in the steady state only the equispaced pair of qubits in the left and right secondary chains have non-zero bi-partite correlations (connected with dashed lines in Fig.\ref{spinchainstopology}(b)). We evaluate the steady state correlator $|\langle\hat{\sigma}^{+}_{n/2+1,\rm{L}}\hat{\sigma}^{-}_{n/2+1,\rm{R}}\rangle|$ and plot it as a function of $n$ in Fig.\ref{diffqubitssetup2}. Our numerical results (dot) confirm that our protocol is capable of generating  bi-partite correlations between the qubits placed at the remote (open) ends of the secondary chains. Moreover, the correlator $|\langle\hat{\sigma}^{+}_{n/2+1,\rm{L}}\hat{\sigma}^{-}_{n/2+1,\rm{R}}\rangle|$ decreases  linearly ($\sim 1/n$) with the increase in the total number of qubits in the chain (solid line). Fig.\ref{diffqubitssetup2} (b) shows that equispaced pair of qubits in the left and right secondary chains have non-zero degree of bi-partite entanglement $\mathcal N_{n/2+1,{\rm L};n/2+1,{\rm R}}$ in the steady state. The resulting impurity of the steady state of the master equation (\ref{open2qubitdep}) is also reflected in the faster drop of pairwise entanglement shown in Fig.\ref{diffqubitssetup2}(b).
  
 \section{Outlook and conclusions} 
  In this work, we have presented a protocol for the generation of  quantum many-body states of coupled qubits arranged in one-dimensional chains. Our scheme makes use of a reservoir engineering strategy to create a maximally entangled state of the two primary qubits. This quantum state is then  transformed to a multi-qubit entangled $W$ state by coherently coupling  the primary chain with secondary chains of qubits. Our work allows a possibility to create genuine many-body entangled states which can be useful for short-distance quantum communication protocols \cite{sbos03,sbos07,akay10,vngo06}, and also for probing some of the fundamental questions related to the emergence of collective phenomena in many-body open quantum systems \cite{sdie08}. Of course, the success of our scheme is critically  limited by energy relaxation  time of the qubits in the primary and secondary chains. However, it might  be possible to use a quantum control technique such as dynamical decoupling strategy  to protect the qubits from the unwanted influence of the environment \cite{gdel10,aber12,nbar13}. 
  
One of the natural candidates to testify our method of generation of many-body quantum states will be systems of trapped atomic ions, which exhibit high degree of controllability  and manipulability \cite{dporr04,kkim11,rbla12}. Trapped atomic ions also provide a flexibility in generating effective qubit-qubit interactions through external lasers. Moreover, the bi-local quantum jump operator that we have considered  has already been implemented in trapped atomic ions \cite{jtba11}. If, on the other hand,  the qubits in our formalism represent actual spin-1/2 particles then $\Delta$ will be the effective `magnetic field' in the $z$ direction and $\theta,\kappa$ will be the XY coupling strengths between neighboring spins. Our coupled qubits arrangements can also be implemented through chains of coupled quantum dots, where each qubit is represented by the presence or absence of a ground-state exciton. In that case, $\Delta$ will be the exciton energy and $\theta,\kappa$ will be the coupling strengths  between neighboring dots. It still remains an open question to address an actual physical implementation of the bi-local quantum jump operator  in a physical setting of coupled quantum dots. 

  There remains an enormous scope for future extensions of our study to explore qubit graphs with more complex geometries \cite{sbos07}. In particular, a natural extension of our present work will be to explore in detail the emergence of many-body quantum states in coupled qubits arranged in quasi-one-dimensional configurations, including  plaquettes and ladders.

\acknowledgments{}
CJ is supported by a York Centre for Quantum Technologies (YCQT) Fellowship. CJ acknowledges fruitful discussions with Erika Andersson, Sougato Bose, Jonas Larson, Patrik \"Ohberg and Tim Spiller. CJ is also thankful to Jonathan Keeling for his help in developing the original TEBD numerical code which is further modified for this paper.

\end{document}